\documentclass[aps,prl,twocolumn, titlepage,showpacs]{revtex4}

\usepackage{graphicx}

\usepackage{dcolumn}

\usepackage{bm}

\begin{document}

\title{Rapid heating and cooling in two-dimensional Yukawa systems}

\author{Yan Feng}
\email{yan-feng@uiowa.edu}
\author{Bin Liu}
\author{J. Goree}
\affiliation{Department of Physics and Astronomy, The University
of Iowa, Iowa City, Iowa 52242, USA}

\date{\today}

\begin{abstract}

Simulations are reported to investigate solid superheating and
liquid supercooling of two-dimensional (2D) systems with a Yukawa
interparticle potential. Motivated by experiments where a dusty
plasma is heated and then cooled suddenly, we track particle
motion using a simulation with Langevin dynamics. Hysteresis is
observed when the temperature is varied rapidly in a heating and
cooling cycle. As in the experiment, transient solid superheating,
but not liquid supercooling, is observed. Solid superheating,
which is characterized by solid structure above the melting point,
is found to be promoted by a higher rate of temperature increase.

\end{abstract}

\narrowtext

\maketitle

\section {I.~Introduction}

Superheated solid is a solid at temperatures above its melting
point~\cite{Bai:05}, and supercooled liquid is a liquid at
temperatures below its melting point~\cite{Ediger:96}. Compared
with supercooled liquid (which is commonly found for many
substances including glasses, colloidal
suspensions~\cite{Weeks:00, Konig:05}, and water~\cite{Smith:99}),
superheated solids are very rare. Indeed, solid superheating was
once thought to be impossible~\cite{Dash:99}, because the thermal
energy will break down the bonds between atoms if the temperature
is higher than the melting point.

In the literature, we find reports of two general methods for
solid superheating. First, if metal or ice is heated with an
ultra-fast heating method~\cite{Herman:92, Iglev:06}, then solid
superheating can occur for a short time. (A limited lifetime of
the superheated solid, before melting, is an indication of what
has been called transient solid superheating~\cite{Cahn:86}.)
Second, solid superheating experiments have been reported for some
specially fabricated samples. For example, if lead is precipitated
into aluminum~\cite{Grabaek:92} or lead layers are sandwiched
between aluminum layers~\cite{Zhang:00}, then the lead can stay in
a solid state at temperatures higher than the melting point of
lead for a long time. (A long lifetime of the superheated solid is
an indication of what has been called metastable solid
superheating~\cite{Cahn:86}.)

In a recent experiment~\cite{Feng:08}, we observed transient solid
superheating in a 2D suspension in a dusty plasma using rapid
laser heating, as reviewed in Sec.~IV. In dusty plasmas, small
particles of solid matter are electrically charged and suspended
in plasmas. Due to their mutual Coulomb repulsion, when confined
electrically these particles can self-organize in a so-called
plasma crystal, in which the particles are arranged in space like
molecules in a crystal or liquid~\cite{Thomas:96, Melzer:96,
Quinn:01, Knapek:07, Zuzic:06, LinI:07}. Since dusty plasmas, like
colloidal suspensions~\cite{Weeks:00, Konig:05}, allow direct
imaging of particles using video microscopy, they allow particle
tracking and measurement of microscopic structure~\cite{Thomas:96,
Melzer:96, Quinn:01, Knapek:07}. This allows direct comparison of
experiment and molecular dynamics simulations~\cite{Knapek:07},
since both the experiment and simulations yield the same
measurable quantities: time series for particle positions and
velocities. Performing a simulation requires specifying the
interparticle potential. One advantage of dusty plasma experiments
with a 2D suspension is that the form of the interparticle
potential is known.  It has been demonstrated experimentally that
the binary interparticle interaction is modeled by a Yukawa
potential~\cite{Konopka:00} when particles are confined to a
single layer that is perpendicular to ion flow, as in our
experiment.

Hysteresis occurs in many physical and other systems. Physical
examples include magnetic~\cite{Norton:03} and electric
polarization~\cite{Priya:02} hysteresis, when time-varying fields
are applied to a condensed matter sample. Hysteresis also occurs
in other fields like biology~\cite{Tyson:01} and
economics~\cite{Baldwin:88}. A hysteresis diagram is typically
prepared by combining time series measurements for two quantities,
for example measurements of magnetization $M$ and applied field
$H$ combined into a single graph with $M$ as the vertical axis and
$H$ as the horizontal axis. For solid superheating experiments,
temperature is the customary horizontal axis~\cite{Grabaek:92}.
For the vertical axis, in our recent experiment~\cite{Feng:08}, we
used the defect fraction, while previous experimenters used some
externally-measured parameter derived typically from an X-ray
diffraction image~\cite{Grabaek:92}.

Here we report results from a molecular dynamics simulation, for
conditions similar to our experiment~\cite{Feng:08}. Molecular
dynamics simulations are idealized models of an experiment because
they use simple physics without some of the peculiarities of the
experiment. We use a 2D Langevin simulation method to model the
entire time series for an experiment, including both heating and
subsequent cooling. (Details are described in Sec.~II.) A previous
paper~\cite{Knapek:07} reported results for a simulation of
cooling.

Our simulation results, Sec.~V, reveal a hysteresis similar to the
one observed in the experiment~\cite{Feng:08}. As in the
experiment, we search for the signatures of solid superheating and
liquid supercooling, and we observe the former but not the latter.
The general agreement between the experiment~\cite{Feng:08} and
the simulation results presented here is helpful in demonstrating
that the experiment observations are merely due to simple physics
processes, and not to peculiarities of the experiment such as
non-Maxwellian velocity distribution or anisotropy~\cite{Feng:08}.
The simulation allows varying some parameters such as friction and
initial defect fraction that are not easily adjustable over a wide
range in the experiment, and this helps in gaining insight into
the experiment. These results may also be helpful in understanding
other 2D experimental systems, including electrons on the surface
of liquid helium~\cite{Grimes:79}, granular fluids~\cite{Reis:06},
and colloidal suspensions~\cite{Konig:05}.

\section {II.~Simulation}

We performed Langevin dynamical simulations. We used a binary
interparticle interaction with a Yukawa pair potential,
\begin{equation}\label{Yukawa}
\phi(r_{i,j})=Q^2(4\pi\epsilon_0r_{i,j})^{-1}exp(-r_{i,j}/\lambda_D),
\end{equation}
where $Q$ is the particle charge, $\lambda_D$ is the screening
length, and $r_{i,j}$ is the distance between particles $i$ and
$j$. Equilibrium Yukawa systems can be classified by the values of
$\Gamma$ and $\kappa$~\cite{Ohta:00, Sanbonmatsu:01}. Here,
\begin{equation}\label{Gamma}
\Gamma=Q^2/(4\pi\epsilon_0ak_BT)
\end{equation}
and $\kappa\equiv a/\lambda_D$, where $T$ is the particle kinetic
temperature, $a\equiv (n\pi)^{-1}$ is the Wigner-Seitz
radius~\cite{Kalman:04}, and $n$ is the areal number density. The
length scale $a$ is related to the lattice constant $b$ (for a
defect-free crystal) by $a = b/1.9046$. Our simulation includes
16~384 particles in a rectangular box of dimensions
$137.5\thinspace b~\times~119.1\thinspace b$. Time scales of
interest are characterized by the inverse of the nominal plasma
frequency,
$\omega_{pd}^{-1}=(Q^2/2\pi\epsilon_0ma^3)^{-1/2}$~\cite{Kalman:04},
where $m$ is the particle mass. (In the experiment~\cite{Feng:08},
$a = 0.45~{\rm{mm}}$ and $\omega_{pd}^{-1} = 30~{\rm{ms}}$.)

We integrate the Langevin equation of motion for each particle.
This equation is
\begin{equation}\label{motion}
m\ddot{\mathbf{r}}_{i}=-\nabla \sum \phi_{ij}-\nu
m\dot{\mathbf{r}}_{i}+\zeta_{i}(t),
\end{equation}
with frictional drag $\nu m\dot{\mathbf{r}}_{i}$ and a random
force $\zeta_{i}(t)$. Particles are allowed to move in a single 2D
plane. Note that we retain the inertial term on the left-hand-side
in Eq.(\ref{motion}), unlike some Brownian-dynamics simulations of
overdamped colloidal suspensions~\cite{Lowen:92}, where it is set
to zero.

The random force, $\zeta_{i}(t)$, is assumed to have a Gaussian
distribution with zero mean. The magnitude of the random force,
characterized by the width of its Gaussian distribution, is chosen
to attempt to achieve a desired target temperature, $T_{ref}$,
according to the fluctuation-dissipation
theorem~\cite{Pathria:1972, Gunsteren:82},
\begin{equation}\label{fluctuation-dissipation}
\langle\zeta_{i}(0)\zeta_{i}(t)\rangle=2m\nu
k_{B}T_{ref}\delta(t),
\end{equation}
where the delta function, $\delta(t)$, indicates that the random
force, $\zeta_{i}(t)$, is local in time.

The fluctuation-dissipation theorem~\cite{Pathria:1972} is useful
for many physical systems, including for example Brownian motion.
It relates dissipation to microscopic fluctuations and the
temperature in thermal equilibrium. This dissipation, which occurs
at a microscopic scale, is of interest also for non-equilibrium
behavior. However, the fluctuation-dissipation theorem does not
accurately model all non-equilibrium systems, for example some
experiments where energy is pumped in~\cite{Lobaskin:06}.
Therefore, we do not expect exact agreement between our
non-equilibrium experiment and a simulation that assumes the
fluctuation-dissipation theorem.

Our simulation mimics our monolayer dusty plasma
experiment~\cite{Feng:08} in the use of a 2D monolayer with a
Yukawa potential with similar values of parameters $\Gamma$ and
$\kappa$, but it differs from the experiment in at least three
ways. First, the heating and friction are explicitly coupled by
Eq.~(\ref{fluctuation-dissipation}), which is slightly different
from the experiment (which is a driven-dissipative
system~\cite{Feng:08, Liu:08}, as described in Sec. IV). Second,
it uses periodic boundary conditions to model an infinite system.
Third, it uses a larger particle number.

The input parameters specified for the simulation include
$\kappa$, $\nu/\omega_{pd}$, and the target $\Gamma_{ref}$
(calculated from a target temperature $T_{ref}$, charge $Q$, and
the Wigner-Seitz radius $a$ using Eq.~(\ref{Gamma})). Here we will
prescribe a waveform for $T_{ref}(t)$ rather than hold it
constant, in order to mimic the rapid heating and cooling in the
experiment. When we change $T_{ref}$, $\nu$ remains constant. This
will cause the random force $\zeta_i(t)$ to become stronger (if
$T_{ref}$ is increased) or weaker (if $T_{ref}$ is decreased)
according to Eq.~(\ref{fluctuation-dissipation}).

We initialized the simulation by starting particles at positions
that would lead to the desired defect fraction, and then running
the simulation for an initialization time $372~\omega_{pd}^{-1}$.
At time $t = 0$, defined as the end of this initialization, we
began recording time series of data for particle positions and
velocities.

The time series we specified for the target temperature,
$T_{ref}$, is presented in Fig.~1. The target temperature was at
first held steady at a baseline (below the melting point) until
about $t=300~\omega_{pd}^{-1}$. Then we began ramping $T_{ref}(t)$
upward at a constant rate, with a rise time of
$10~\omega_{pd}^{-1}$ to a maximum value of $T_{ref}$, which we
specified well above the melting point. This procedure was
intended to mimic the rapid heating in the experiment. Next, we
held $T_{ref}(t)$ constant for a duration of
$1806~\omega_{pd}^{-1}$, nearly matching the duration of steady
laser heating in the experiment. Then, we ramped $T_{ref}(t)$ back
down to its baseline in a fall time of $30~\omega_{pd}^{-1}$, to
mimic the rapid cooling of the experiment. Finally, we held
$T_{ref}(t)$ constant, recording data until
$t=3325~\omega_{pd}^{-1}$. This final stage corresponds to the
long period of recrystallization observed in the
experiment~\cite{Feng:08}.

Here we review other details of our simulation. We used the
Langevin integrator of Gunsteren and
Berendsen~\cite{Gunsteren:82}. A time step of
$0.037~\omega_{pd}^{-1}$ and periodic boundary conditions were
used. We truncated the Yukawa potential at radii beyond 12~$a$,
with a switching function to give a smooth cutoff between 12~$a$
to 14~$a$ to avoid an unphysically sudden force when a particle
moves a small distance.

We performed two simulations. Run~1 was intended to approximate
the friction in the experiment, $\nu/\omega_{pd}=0.066$,
corresponding to $\nu=2.2~{\rm{s^{-1}}}$ and
$\omega_{pd}=33.3~{\rm{s^{-1}}}$. Run~2 had a ten-fold higher
friction, $\nu/\omega_{pd}=0.66$, corresponding to
$\nu=22~{\rm{s^{-1}}}$. Both of these runs began with particles
arranged in a solid structure having an initial defect fraction
(concentration) of $0.027$ similar to the experiment.

As a test, we repeated the simulations reported here with a
different initial condition of a defect-free crystal. We found
that the results are similar enough that our conclusions are
unaffected by the initial defect fraction.

The time series for temperature in the experiment and simulations
are presented in Fig.~1. In both cases, the observed temperature
$T$ is calculated from the measured values of the mean-square
velocity fluctuations. The observed temperature time series from
the simulation is similar to the temperature time series from our
experiment for Run~2, with the high friction. But at a lower
friction in Run~1, the temperature changes more slowly. As
compared to Run~2, the slow rate of temperature change in Run~1 is
due to a smaller random force $\zeta_i(t)$ from
Eq.~(\ref{fluctuation-dissipation}) when $T_{ref}$ is changed.

The difference in the rate of change of temperature, for the
simulation as compared to the experiment, is attributed to the
different ways that heating and friction are related in the
experiment (where they are independent) and simulation (where they
are explicitly coupled through Eq.(\ref{fluctuation-dissipation}),
as discussed above). Because of this, it is difficult to match
both the observed temperature time series $T(t)$ and gas friction
$\nu$ in simulations. Our simple waveform for the target
temperature, $T_{ref}(t)$, allows us to match $T(t)$ or $\nu$, but
not both. Therefore, we will compare results for two cases: Run~1
where we match the friction $\nu$, and Run~2 where we nearly match
the time series for observed temperature $T(t)$. Comparing these
two runs will be useful in assessing the relative importance of
friction and rate of temperature change for solid superheating.

\section {III.~Diagnostics}

Here we introduce the diagnostics used to test for solid
superheating and liquid supercooling. Our main result will be time
series for two variables, which we will combine to construct a
hysteresis diagram. One of these variables will be the observed
temperature, $T$. The other variable will be chosen from three
structure indicators, which are calculated from the particle
positions. First, we identify defects and calculate defect area
fraction by calculating Voronoi diagrams~\cite{Quinn:01}. Second,
we measure the short-range translational order using the height of
the first peak of the pair correlation function
$g(r)$~\cite{Nosenko:06}. Third, we measure the short-range
orientational order using the bond-angular-order parameter,
$G_\theta$~\cite{Schweigert:99}. We present a detailed explanation
of these three structure indicators next.

A Voronoi diagram is calculated from particle positions of each
frame~\cite{Quinn:01}. Figure~2 shows the Voronoi diagram
calculated for Run~1 before rapid heating. In the case of a
defect-free 2D crystal, the Voronoi diagram would include only
six-sided polygons. When defects are present, they are identified
by the presence of non-six-sided polygons, as in Fig.~2, where the
number of sides is indicated by different colors. To reduce the
information in a Voronoi diagram to a single parameter, we
calculate the defect fraction as the ratio of the areas of all
non-six-sided polygons to the area of the entire Voronoi diagram.
The defect fraction can vary from zero for a defect-free crystal
to roughly $0.3$ for a liquid.

A feature that can be identified easily by examining Voronoi
diagrams is the presence of different domains that collectively
form a polycrystalline solid. For Run~1, in Fig. 2, we see that
most defects are not distributed sparsely, but instead tend to
self-organize by forming strings that serve as domain walls. In
each domain, there is a crystalline region that has an angular
orientation that is different from the next. Domains divided by
domain walls are features also found in our
experiment~\cite{Feng:08}.

The pair correlation function $g(r)$ is calculated from particle
positions; and it can be reduced to a single parameter by
measuring the height of its first peak. This height serves as a
measure of the local translational order, and it can vary upward
from roughly 2 for a liquid to arbitrarily large values for a
solid, depending on temperature and defects. This parameter
generally does not exhibit significant jumps as disorder
increases, so that we do not use it to distinguish liquid from
solid.

The bond-angular-order parameter $G_\theta$~\cite{Schweigert:99}
is calculated from angles between nearby particles. The value of
$G_\theta$ varies from zero for a gas to unity for a defect-free
crystal. For a solid, $G_\theta$ is less than unity if there are
defects. The calculation of this parameter, like the defect
fraction and the height of the first peak of $g(r)$ listed above,
involves an average over a sample area, which in the case of our
simulation is the entire simulation box. In previous simulations
with slowly-varying temperature~\cite{Schweigert:99}, it was found
that $G_\theta$ served as a useful indicator of melting because of
a distinctive jump in its value. This jump occurs at $G_\theta =
0.45$ for different 2D physical systems, including a 2D Yukawa
system, which is useful for quantifying the melting point.
However, we find that $G_\theta$ is very sensitive to the presence
of domains within the sample area, because terms entering the
calculation of $G_\theta$ for one domain can cancel those from
another domain, so that the value of $G_\theta$ depends on the
size of the sample area and how many domains it includes. The
larger the number of domains enclosed, the smaller the value of
$G_\theta$. Therefore, $G_\theta$ will be more useful when melting
a defect-free crystal with only a single domain than for melting a
polycrystalline solid like the one in Fig.~2 and in the
experiment~\cite{Feng:08}.

Of the three structure indicators listed above, we choose the
defect fraction as the variable to present in the vertical axis of
the hysteresis diagram. This choice has the advantage that, unlike
$G_\theta$, it is not highly sensitive to the size of the sample
area, and it has a slower response to a change in temperature than
the height of the first peak of $g(r)$~\cite{Feng:08}. Our
hysteresis diagram will therefore have defect fraction and
temperature as the vertical and horizontal axes, respectively.

The type of hysteresis that is observed here is rate dependent.
Consequently, the general appearance of a hysteresis diagram will
depend on how rapidly the temperature is varied. We illustrate
this in the sketch in Fig.~3. If temperature is varied rapidly,
the hysteresis will be most extreme, while if it is varied
infinitely slowly in a quasistatic process, hysteresis will vanish
and the curve will retrace itself exactly when melting and
solidifying.

The signature of solid superheating or liquid supercooling can be
easily identified in a hysteresis diagram~\cite{Feng:08}. A
horizontal row of data points across the melting point means that
the temperature has changed across the melting point while the
structure has not changed yet. This is sketched at the bottom and
top of Fig.~3.

To determine the melting point, we rely on the phase diagram, Fig.
6 in Ref.~\cite{Hartmann:05}, for a 2D equilibrium Yukawa system.
This phase diagram provides a curve in the $\Gamma$ - $\kappa$
parameter space. Using this curve is straightforward for our
simulation because we specify $\kappa$, so that the curve directly
yields the $\Gamma$ (and therefore the temperature) for the
melting point. We also use the same curve to determine the melting
point for the experiment using the same procedure and an
experimentally-measured value for $\kappa$.

\section {IV.~Review of Experiment}

Here we review the experiment reported in Ref.~\cite{Feng:08} and
provide further discussion of its physics. A single horizontal
layer of electrically-charged polymer microspheres was
electrically levitated in a glow-discharge plasma, forming what is
called a dusty plasma. Viewing the suspension from above with a
video camera, movies of particle motion were recorded. Initially,
particles were self-organized in a nearly crystalline solid
lattice. Particle motion was cooled by friction on the ambient
rarefied neutral gas. Later, an external source of heating was
applied suddenly. This heating source was a cw laser beam,
rastered in a Lissajous pattern to give particles kicks at nearly
random times~\cite{Liu:08, Nosenko:06}. In steady state, the
particle kinetic temperature is determined by a balance of
external laser heating and frictional gas drag
cooling~\cite{Feng:08}. After applying the external heating for
about 55~s, it was suddenly stopped. During the initial phase of
external heating, the temperature increased rapidly, at about
$20~000~{\rm{K/s}}$. After a delay of about 0.25~s the suspension
melted, as judged by a change in defect fraction.

This delay was interpreted as an indication of solid superheating.
An additional indication is the signature of solid superheating
that can be identified in the experimental hysteresis diagram,
Fig.~4, as a horizontal row of about 15 data points. This row of
data points begins at about the melting point, and continues to
about 10~000~K, well above the melting point.

The solid superheating had a limited duration, which we interpret
as an indication that it is a kind of transient solid
superheating.  In general, one could identify solid superheating
as being either transient or metastable, depending on the duration
of the solid structure after increasing the temperature above the
melting point. The distinction between transient and metastable
superheating has been previously mentioned in a review of the
literature~\cite{Cahn:86}. For our experiment, we judge the
duration of the superheated solid by comparing its lifetime of
about 0.25~s to another important time scale for particle motion:
the period of oscillation corresponding to the Einstein frequency,
$\omega_E$. Here $\omega_E$ has the usual meaning: it is the
oscillation frequency that a charged particle's motion would have
in a cage formed by all the other particles, if all the other
particles were stationary. The Einstein frequency for our
experiment can be estimated from a combination of our experimental
measurement of the plasma frequency, $\omega_{pd} =
33.3~{\rm{s^{-1}}}$, and a previous simulation that provided a
relationship between $\omega_{pd}$ and
$\omega_{E}$~\cite{Kalman:04}. This yields an estimate for in our
experiment of $\omega_E = 0.612\thinspace\omega_{pd} =
20.4~{\rm{s^{-1}}}$. The corresponding period of oscillation for
the charged particle in the experiment is $\tau_E = 2\pi/\omega_E
= 0.31~{\rm{s}}$. Comparing now to the experimentally observed
lifetime of about 0.25~s for the superheated solid, we find that
the lifetime was only about one oscillation period, before melting
occurred. Therefore, we interpret our experimental results as an
indication of transient, not metastable superheating.

The underlying reason for the solid superheating in the experiment
is simple to understand, now that the time scales have been
determined. Initially, in the solid below the melting point,
particles are caged by their nearest neighbors. Caged particle
motion in a solid consists mainly of oscillations, with a turning
point located well within the cage. In a full period of
oscillation, characterized by a $\tau_E$, a particle's trajectory
has two turning points. As rapid heating is suddenly applied,
particles in the cage are accelerated, the cage distorts as other
particles are also accelerated, and the enclosed particle can
eventually decage and thereby generate a defect. In the
experiment, the time indicated by the hysteresis diagram for this
decaging to occur is about 0.25~s, about the same as $\tau_E =
0.31~{\rm{s}}$. Comparing these two values indicates that after
sudden heating is applied, a particle typically decages after
bouncing about twice in the cage. This short-lived stage of
bouncing about twice before decaging corresponds to the transient
superheated solid.

During the experiment, the single-layer particle suspension was
not constrained in its size. In principle, its areal number
density could vary in time. We calculated a time series for the
areal number density, and we found that there was no significant
expansion as the temperature increased. The areal number density
remained constant within 1.5\% during the
experiment~\cite{Feng:08}, despite very large temperature changes
of an order of magnitude. It is interesting that despite the
extreme softness of this suspension, its volume varies so little
with temperature.

Another major result from our experiment was that the signature of
liquid supercooling was lacking in the hysteresis diagram, Fig.~4.
A horizontal row of data points extending below the melting point
is absent in this hysteresis. Instead, the defect fraction drops
dramatically as the temperature decreases. In Sec. V, for the
simulation results, we will examine the hysteresis diagrams to
determine whether the same signatures of solid superheating and
liquid supercooling are present.

One feature of the hysteresis diagram that requires explanation is
the gap in data points at the lower left of Fig.~4. This gap is
due to the finite data-recording time in the experiment. After the
initial rapid cooling, a very slow recrystallization takes place.
During the recrystallization, crystalline domains gradually grow
in size by merging with neighboring domains. The merging process
is slow because a domain must rotate until its orientation aligned
with a neighboring domain. This process becomes increasingly slow
as the remaining domains become larger, as can be seen in the
Voronoi movie from the experiment~\cite{Feng:08}. The camera in
the experiment had a finite memory, allowing the recording of a
movie limited to 100~s duration for the entire experiment. A
similar gap will occur in the hysteresis diagram for the
simulation data, Sec. V, because of the expense of running the
simulation to the completion of the same slow recrystallization
process.

Previous to our experiment~\cite{Feng:08}, Knapek {\it et al.}
reported another experiment~\cite{Knapek:07} to study the
recrystallization during cooling. They used a similar dusty plasma
with a single-layer suspension of microspheres. They heated their
suspension suddenly by applying an electrical pulse to wires.
Using video microscopy, particle motion was recorded well after
the pulse was completed, so that the experimenters observed the
cooling process, but not the heating process. As in our
experiment~\cite{Feng:08}, this cooling process included a rapid
cooling followed by a slow recrystallization. Like us, they
reported time series for temperature and defect fraction; they
also reported correlation lengths as measures of orientational and
translational order, serving roles similar to $G_{\theta}$ and
height of the first peak of $g(r)$. They found that temperature
decreases more rapidly than defect fraction~\cite{Knapek:07}, a
result that we verified in ~\cite{Feng:08}. They also found that
orientational order drops much more slowly than translational
order, and attributed this to the presence of domains in various
orientations during the slow recrystallization
process~\cite{Knapek:07}. Our experiment differed by using laser
heating, which did not disturb the particle layer severely.
Because of this, we were able to record particle motion during
both heating and cooling, allowing us to prepare hysteresis
diagrams.

\section {V.~Simulation~Results}

In our Langevin dynamical simulation, we found a hysteresis, as in
the experiment~\cite{Feng:08}. The hysteresis diagrams were
prepared by combining time series for observed temperature,
Fig.~1(b), and defect fraction. The data in these time series were
recorded at time intervals  $0.37~\omega_{pd}^{-1}$. Numerical
noise in the simulations was reduced below the level in the
experiment by using a large number 16 384 particles in the
simulation, about 15 times larger than in the experiment. As a
result, the hysteresis curve is less noisy for the simulation than
for the experiment. The two runs described below began with
different initial particle positions, but the same defect
fraction.

Recall that when we change $T_{ref}$ in the simulation, the
friction $\nu$ remains constant, and the magnitude of the random
force $\zeta_i(t)$ is changed according to
Eq.~(\ref{fluctuation-dissipation}). The observed temperature,
$T(t)$, will lag the target temperature, $T_{ref}(t)$, because of
the time required for the random force to accelerate particles.
This lag in the temperature change is seen in Fig.~1, especially
for the low friction case, Run~1.

\subsection {A.~Run~1 - low friction}

For Run~1 we found the hysteresis diagram, Fig.~5(a). For this
run, the friction was as low as in the experiment, but the
observed temperature changed at a slower rate.

The signature of solid superheating in the simulation for Run~1 is
not as clear as in the experiment. Examining the bottom of the
hysteresis diagram, Fig.~5(a), we observe that the defect fraction
begins to increase noticeably before the temperature has exceeded
the melting point. The structure, as measured by defect fraction,
is no longer the same as it was in the initial solid, although it
more nearly resembles a solid than a liquid. A rapid increase in
defect fraction ensues at temperatures somewhat higher than the
melting point.

The weaker signature of solid superheating in the simulation might
be due to the different temperature time series, as compared to
the experiment. As shown in Fig.~1, the time series for observed
temperature in experiment and Run~1 do not match exactly. Since
the hysteresis is rate dependent, a slower change in temperature
will tend to lack a signature of solid superheating, as sketched
in Fig.~3.

The signature of liquid supercooling in Run~1 is lacking, as it
was in our experiment. Instead of remaining constant as the
temperature decreases as would be required for supercooling, the
defect fraction drops dramatically.

\subsection {B.~Run~2 - higher friction}

For Run~2 we found the hysteresis diagram, Fig.~5(b). For this
run, the friction was ten times higher than in the experiment, but
the observed temperature changed at about the same rate.

The signature of solid superheating signature for Run~2 resembles
the experiment more nearly than for Run~1.  This leads us to
conclude that a high rate of temperature change, as in Run~2, is
important for attaining solid superheating.

The signature of liquid supercooling remains lacking in Run~2.
This result for both simulation runs and the experiment suggests
that liquid supercooling is not easily attained in this physical
system, for the rate of temperature change that we explored here.

\section {VI.~Discussion}

We have simulated our rapid heating and cooling
experiment~\cite{Feng:08}. We used a Langevin simulation of a 2D
Yukawa system, with a temperature that was ramped in time by
specifying a target temperature. By combining time series for
observed values of temperature and defect fraction, we produced
hysteresis diagrams. These diagrams allow an inspection for the
signatures of solid superheating and liquid supercooling.

We draw three chief conclusions. First, the simulations are
capable of producing hysteresis as in the experiment. The physics
incorporated in the simulation is very simple, in comparison to
the experiment which has more complications. Our finding that
hysteresis arises in both the simulation and experiment indicates
that the cause of the hysteresis is simple physics, and not a
peculiarity of the experiment.

Second, in both the experiment and simulation, the signature of
liquid supercooling was lacking. This result is of interest
because it is an open question whether there can be any
one-component 2D systems that behave like a supercooled
liquid~\cite{Konig:05}.

Third, we found that the hysteresis curve for simulation and
experiment most nearly agree when the rate of temperature change
is matched. Our simulation method, together with our choice of a
time series for the target temperature $T_{ref}(t)$, allowed us to
match either the time series for observed temperature $T(t)$ or
the friction $\nu$ for the experiment and simulation, but not
both. We found that the hysteresis in the experiment was most
nearly duplicated in the simulation run with the same rapid change
of the observed temperature. This result leads us to conclude that
a high rate of temperature change is an important requirement for
attaining solid superheating.

We thank F. Skiff for helpful discussions. This work was supported
by NASA and DOE.

\begin{figure}[p]
\caption{\label{temperature} (color online). Time series for
temperature. (a) For the experiment, the observed temperature was
calculated from the mean square velocity fluctuation. (b) For the
simulation, the target temperature time series (red solid line)
was prescribed. The resulting observed temperature, computed from
mean square velocity fluctuations, are shown for Run 1 with low
friction (blue dash line) and Run 2 with high friction (green dot
line).}
\end{figure}

\begin{figure}[p]
\caption{\label{Voronoi} (color online). (a) Voronoi diagram
calculated from our simulation data, before rapid heating when
$\Gamma > 400$. (b) Magnified view of the portion inside the green
square in (a). For both panels, the horizontal and vertical
dimensions are normalized by the lattice constant $b$. For both
Run~1, shown here, and Run~2, the initial conditions had the same
defect fraction 0.027. Note that at this level of defect fraction,
defects are arranged mainly in strings, forming domain walls that
separate domains with different orientations. Due to the presence
of these domains, $G_{\theta}$ is only 0.216.}
\end{figure}

\begin{figure}[p]
\caption{\label{sketchmap} (color online). Sketch of a hysteresis
diagram for a solid-liquid transition. The vertical axis, defect
fraction, is a structure indicator. Hysteresis, if it occurs, may
depend on the rate of temperature change. It is not expected if
temperature is changed infinitely slowly in a quasistatic
process.}
\end{figure}

\begin{figure}[p]
\caption{\label{expsimulation} Hysteresis diagram for the
experiment. Time series for observed temperature $T$ and defect
fraction, both recorded at time intervals of 18 ms, were combined
to produce this diagram. At the bottom, the horizontal row of data
points crossing the melting point was interpreted
in~\cite{Feng:08} as showing the signature of solid superheating.
Reprinted from~\cite{Feng:08}.}
\end{figure}

\begin{figure}[p]
\caption{\label{twosimulation} Hysteresis diagrams for (a) Run~1
and (b) Run~2. In both runs, the same initial defect fraction and
target temperature time series were used, but in Run~1 a lower
friction resulted in a slower rate of temperature change.  These
diagrams were made by combining time series for defect fraction
and observed temperature $T$, recorded at a time interval of
$0.37~\omega_{pd}^{-1}$.}
\end{figure}

\end{document}